\documentclass[twocolumn]{aa}
\usepackage{graphicx}
\usepackage{natbib}
\bibpunct{(}{)}{;}{a}{}{,} 
\begin{document}
 \title{TNG Near-IR Photometry of five Galactic Globular Clusters\thanks{Based on observations made with the Italian
Telescopio Nazionale Galileo (TNG) operated on the island of La Palma by the Centro Galileo Galilei of the Consorzio
Nazionale per l'Astronomia e l'Astrofisica (CNAA) at the Spanish Observatory del Roque de los Muchachos of the Instituto
de Astrofisica de Canarias.}}
%

   \author{E. Valenti\inst{1,}\inst{2},
	 F.~R.~Ferraro\inst{1},
	 S. Perina\inst{1},
	 L. Origlia\inst{2}
	 }

   \offprints{E. Valenti}

   \institute{Dipartimento di Astronomia Universit\`a 
              di Bologna, via Ranzani 1, I--40127 Bologna, Italy \\
              \email{elena.valenti2@studio.unibo.it; ferraro@bo.astro.it}
	     \and
	         INAF--Osservatorio Astronomico di Bologna, via Ranzani 1,
                 I--40127 Bologna, Italy \\
              \email{origlia@bo.astro.it}
             }

   \date{}

   \abstract{We present near--infrared J and K observations of giant stars in five metal\--poor Galactic Globular
Clusters (namely M3, M5, M10, M13 and M92) obtained at the Telescopio Nazionale Galileo (TNG).
This database has been used to determine the main photometric properties of the red giant branch (RGB) from the
(K,J-K) and, once combined with the optical data, in the (K,V-K) Color Magnitude Diagrams.
A set of photometric indices (the RGB colors at fixed magnitudes) and the major RGB evolutionary features (slope, bump,
 tip) have been measured. The results have been compared with the relations obtained by \citet{F00} and
with the theoretical expectations, showing a very good agreement.
 
   }

 \authorrunning{Valenti et al.}
 \titlerunning{}
   \maketitle
%

\section{Introduction}
This paper is part of a long-term project devoted to the detailed study of the 
Red Giant Branch (RGB) properties of
Galactic Globular clusters (GGCs) using the Color--Magnitude Diagrams (CMDs) and 
Luminosity Functions (LFs) in the near--infrared (IR) \citep[see e.g.][hereafter F00]{F00}.
 
The near--IR spectral range is particularly suitable to study the cool
stellar sequences, being the most sensitive to low temperatures.
In addition, compared to the visual range,
the reddening corrections and the background
contamination by Main Sequence (MS) stars are much less severe, allowing to
properly characterize the RGB even in the innermost core region 
of stellar clusters, affected by severe crowding.
This is well known for two decades, and  
several authors have used IR photometry  
to derive the main RGB properties,
starting from the pioneering survey of $\sim$30 GGCs by \citet{FCP83}, 
based on single--channel detectors and aperture photometry.

In the 90's with the advent of bidimensional IR array detectors 
with pixel sizes and overall performances similar
to those of optical CCDs, 
larger and more complete samples of Population~II RGB stars with high 
photometric accuracy have been provided \citep[see e.g.][]
{DS91,DS94a,DS94b,F94,Mi95,Mis95,F95,M95,K95,KF95,Gu98,F00,Yaz03}.

By combining near--IR and optical photometry one can also calibrate a few major 
indices with a wide spectral baseline, which turn out to be very sensitive to the stellar parameters, 
like for example the (V-K) color index, possibly the best photometric thermometer 
for cool giants. 
F00 presented high-quality near-IR CMDs of 10 GGCs
spanning a wide metallicity range which have been used $(i)$ to calibrate several observables describing the
RGB physical and chemical properties; $(ii)$ to detect the major RGB evolutionary features as the RGB bump and the
RGB tip; $(iii)$ to prepare followup observations devoted to better understand the mass loss process 
\citep[see e.g.][]{O02}.
In this paper we present an extension of the F00 work to five additional metal--poor GGCs.

The observations and data reduction are presented in \S 2, while \S 3 describes the 
properties of the observed CMDs. \S 4 is devoted to 
derive the mean RGB features from the combined study of CMDs and LFs. We summarize
our conclusions in \S 5.

\begin{figure*}
\centering
\includegraphics[width=12cm]{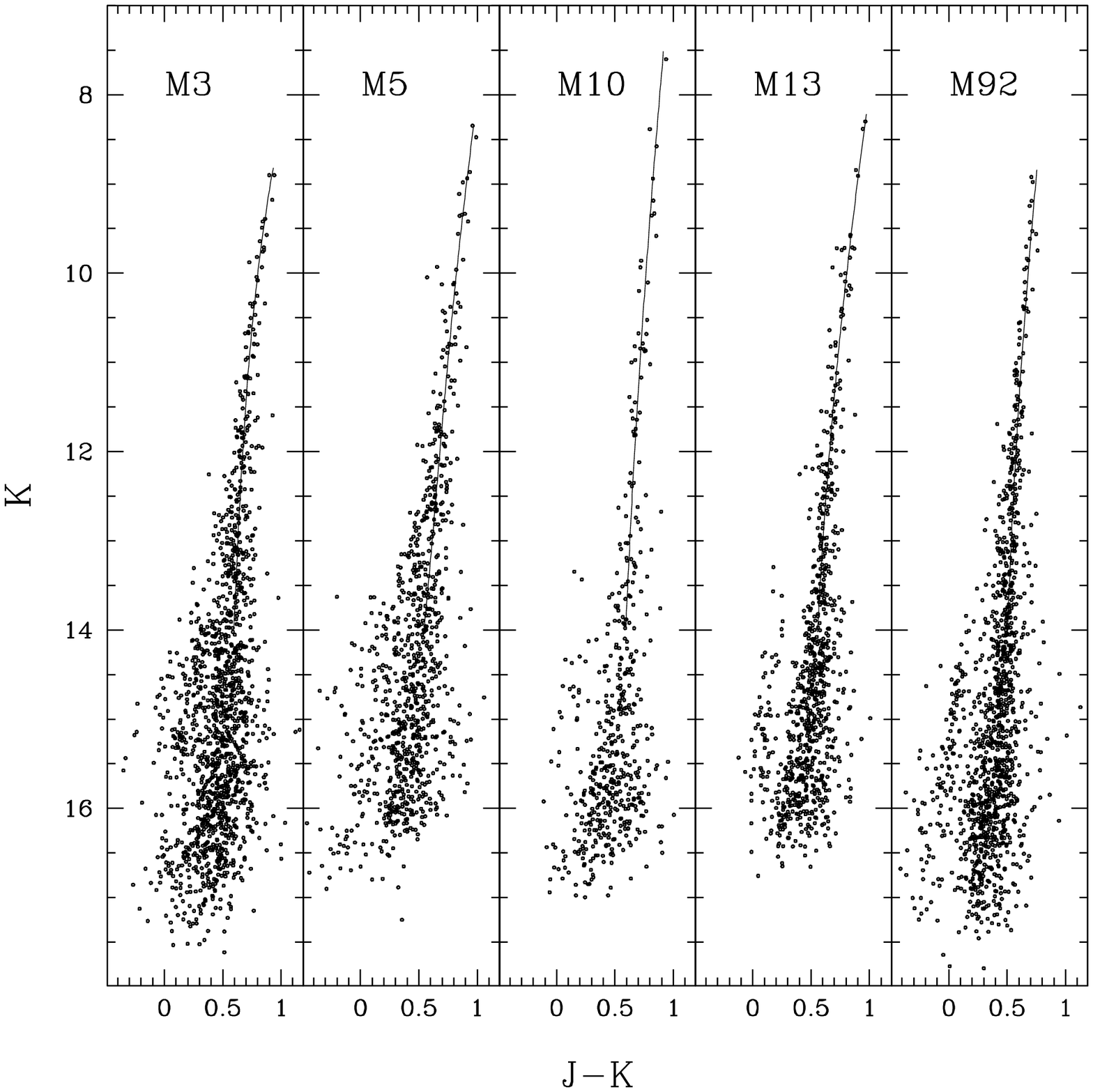} 
\includegraphics[width=12cm]{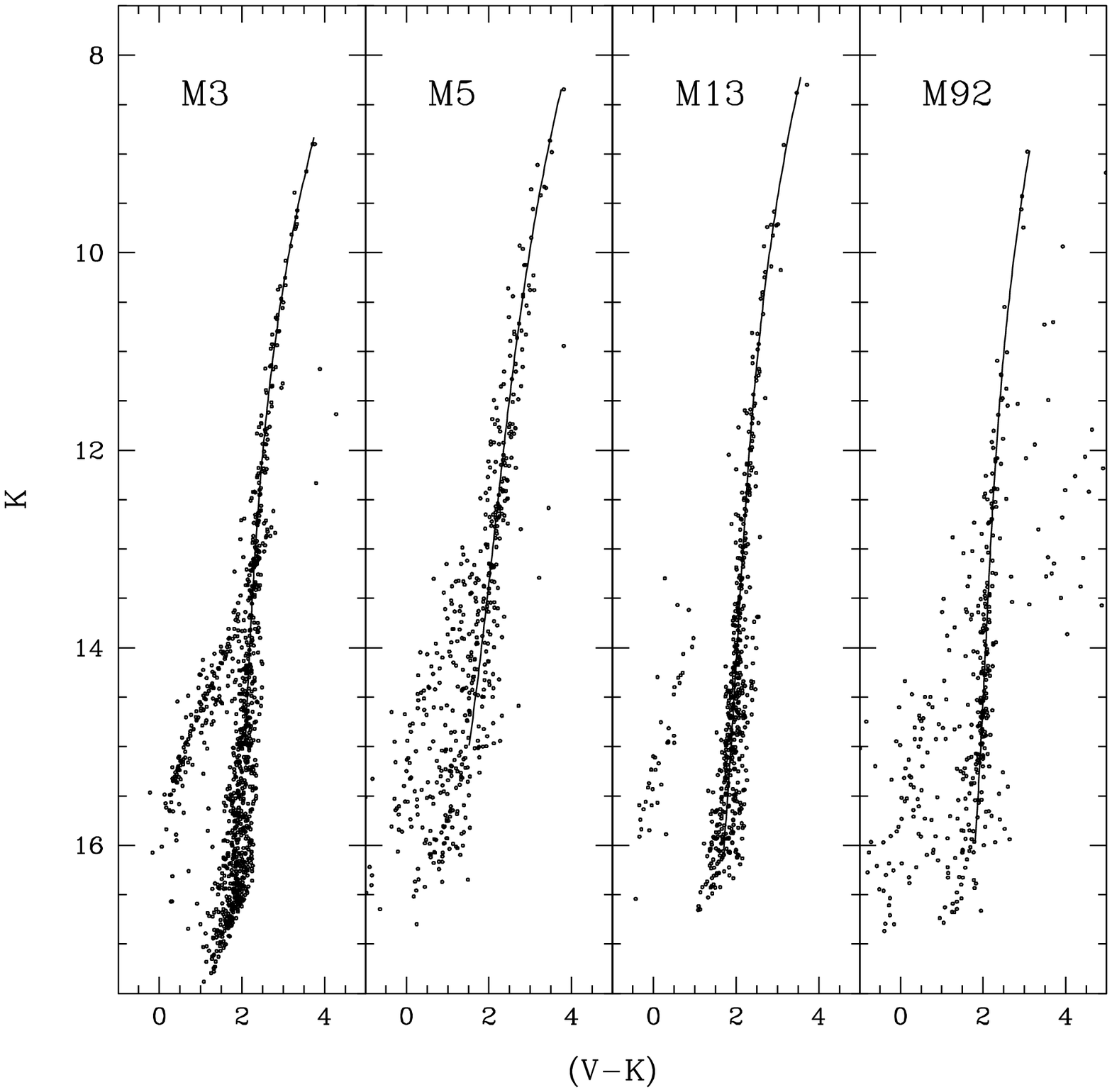} 
\caption{K,J-K (upper panels) and K,V-K (lower panels) CMDs for the observed clusters.
The thick line in each panel indicates the fiducial ridge line of the RGB.}
\label{cmd}
\end{figure*}

\section{Observations and data reduction}

A set of J and K images were secured at the Telescopio Nazionale Galileo (TNG),
in the Canary Islands during 3 nights on May 10,11 and 13th, 2000, 
using the near-IR camera ARNICA equipped with a
NICMOS-3 $256\times256$ array detector. By using a magnification of 0.35 $''/px$, for a total field of view 
 of $1.5'\times1.5'$, 5 GGCs, namely M3, M5, M10, M13 and M92,
have been observed. Table \ref{par} lists the main parameters adopted for the program clusters: 
 the metallicity in the original scale of \citet{Z85}, in the \citet{CG97} (hereafter CG97) scale, and the
global scale as defined by \citet{F99}, the reddening and the distance modulus from Table 1 of F99.
 The central
region of the cluster was mapped in all the targets with the exception of M10 for which
only a field at ${\sim}1$' East from the cluster center was secured.
In M92, besides the central field, an additional partially overlapping field has been observed at ${\sim}1.5$'
South--East from the cluster center.

During the observations the average seeing  was $ 0.8''-1''$.
Each acquired J and K image was the average of 120 single exposures of 1-s detector integration
time (DIT). A series of typically 4 in J and 8--12 images in K were secured on each target, for a total integration time of
${\approx}$ 8 and 24 minutes in J and K, respectively. 
The sky
was measured at a distance of ${\sim}$8'--10' from the cluster center. More details on the pre--reduction
procedure can be found in \citet{F94}, \citet{M95}. For each field we combined all the available sky--corrected
images, in order to obtain a median--averaged image with a signal to noise ratio ${\geq}50$.

The photometric reduction was carried out by using the ROMAFOT package \citep{B83,BI89}. The Point Spread Function
(PSF) fitting procedure was adopted to determine the instrumental magnitude of the stars in the frame.
 The full description of the reduction
procedure is shown in previous papers \citep[see][]{F94,F97} and it will not be repeated here. We just report
that the automatic search for the object detection was performed (in each filter) in the median--averaged image. 
As usual, the
mask with the object positions was used as input for the PSF--fitting procedure and
a catalog listing the instrumental magnitudes measured in each image 
was obtained. Magnitudes in each filter were then transformed to a homogeneous photometric system.
A final catalog with the average instrumental J and K magnitudes from different images of the same field was finally
obtained.

Since the observations were performed under not perfect photometric conditions, we adopted the following
procedure to calibrate the data--set:\\
{\it i}) The instrumental magnitudes were first transformed into the {\it Two Micron All Sky Survey} (2MASS) 
photometric system{\footnote{In doing this we used the Second Incremental Release Point Source Catalog of 2MASS}}. 
Only zero--order polynomial transformations were used.\\

{\it ii}) The catalogs in the 2MASS photometric system were finally transformed into the F00 system by
using the following transformations:\\

$J_{F00}=J_{2MASS}+0.06$ \\

$K_{F00}=K_{2MASS}+0.05$ \\

The transformations were empirically calibrated by using more then 5000 stars in common between the 2MASS catalogs
and the catalogs presented by F00 for 10 GGCs.

Fig.~\ref{cmd} present the K,J-K CMDs for the observed clusters calibrated in the F00 system.
An overall uncertainty of ${\pm}0.05$ mag in the zero point calibration, both in the J
and K bands, has been estimated{\footnote {The photometric catalogs, 
calibrated in the F00 system, are availables in the electronic form.}}.

\begin{table}
\begin{center}
\caption[]{Adopted parameters for the observed clusters.}
\label{par}
{\tiny
\begin{tabular}{lccccc}
\hline\hline
\\
Name & $[Fe/H]_{Z85} $&$[Fe/H]_{CG97}$& $[M/H]$  & $E(B-V)$ &  $(m-M)_0$ \\
\\
\hline
\\
M3  &-1.66& -1.34 & -1.16 &  0.01 & 15.03 \\
M5 &-1.40&-1.11 & -0.90 & 0.03 & 14.37 \\
M10 &-1.60&-1.41 & -1.25 & 0.28 & 13.38\\
M13 &-1.65& -1.39 & -1.18 & 0.02 &14.43 \\
M92 &-2.24&-2.16 & -1.95 &0.02 & 14.78\\
\\
\hline
\end{tabular}
}
\end{center}
\end{table}
 
\begin{figure}
\centering
\includegraphics[width=8.7cm]{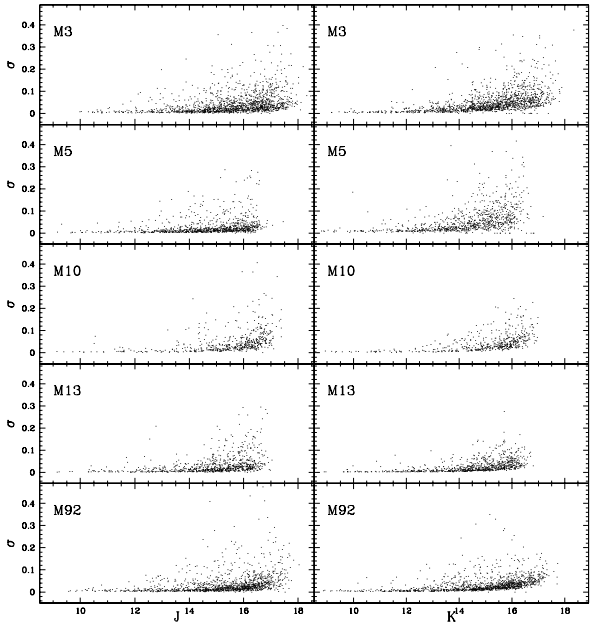} 
\caption{Internal photometric errors for stars detected in the observed cluster sample. The {\it rms} values of the
frame\--to\--frame scatter expressed in magnitudes are plotted as a function of the J (left panels) and K (right panels)
calibrated magnitudes.}
\label{rms}
\end{figure}

\section{Color Magnitude Diagrams}

More then 4500 stars are plotted in the (K,J--K) and (K,V--K) CMDs shown in Fig.~\ref{cmd}.
Since the IR observations mapped the central regions of the clusters, we mostly used optical photometry from
HST (M3: \citet{F97}, \citet{Ro99}; M5: \citet{San96}; M13: \citet{F97b}; M92: Ferraro et al.(2003) 
in preparation).

The main features of the CMDs presented in Fig.~\ref{cmd} can be schematically summarized as follow:\\
{\it i}) The RGB is quite well populated and allows us a clean definition of the mean ridge line in all the program
clusters. The observations reach $K{\approx} 17$ mag and are deep enough to detect the base of the RGB at
${\Delta}K{\approx}7-8$ mag fainter than the RGB tip.\\
{\it ii}) In the combined CMDs the Horizontal Branch (HB) stars are clearly separable 
from the RGB stars. The HB appears as a sequence
which has an almost vertical structure in all the CMDs. This is not surprising since intermediate metal--poor
clusters are expected to have blue HB.\\
{\it iii}) The RGB is well populated in all program clusters, even in the
brightest magnitude bin, with the possible exception of M10.
 The large size of the available sample allows a meanful estimate of the major evolutionary
features along the RGB, namely the RGB bump and tip.

\subsection{Internal Photometric errors}

A reasonable estimate of the internal photometric accuracy of the IR photometry presented here can be estimated 
from the {\it rms} frame\--to\--frame 
scatter of  multiple stars measurements. 
Fig.~\ref{rms} shows the distribution of the obtained values \citep[see][]{F91},
 for all the stars detected in each program cluster and in each band as a function of the final
calibrated magnitude.   
As expected the internal errors significantly increase at fainter magnitudes due to photon
statistics. In addition to this general trend, a number of stars show a ${\sigma}$ larger than the mean value:
these stars are either variable stars (mainly RR Lyrae) observed at random phase (M~3, M~5 and M~92 harbor a large
population of such variables) or stars lying in the innermost region of the clusters severely affected by crowding
conditions. Nevertheless, over most of the RGB extension, the internal errors are quite low
(${\sigma}_K{\sim}{\sigma}_J<0.03$ mag), increasing up to ${\sigma}_K{\sim}{\sigma}_J{\sim}0.06$ at $K{\geq}16$.
             
\subsection{Comparison with previous photometry}

The clusters presented here have been subject of several photometric and spectroscopic investigations in the optical,
particularly M3 and M13, since they are a classical HB {\it Second Parameter pair}. 
However, only a few papers in the literature
present IR photometry for these clusters.

\begin{figure}
\centering
\includegraphics[width=8.7cm]{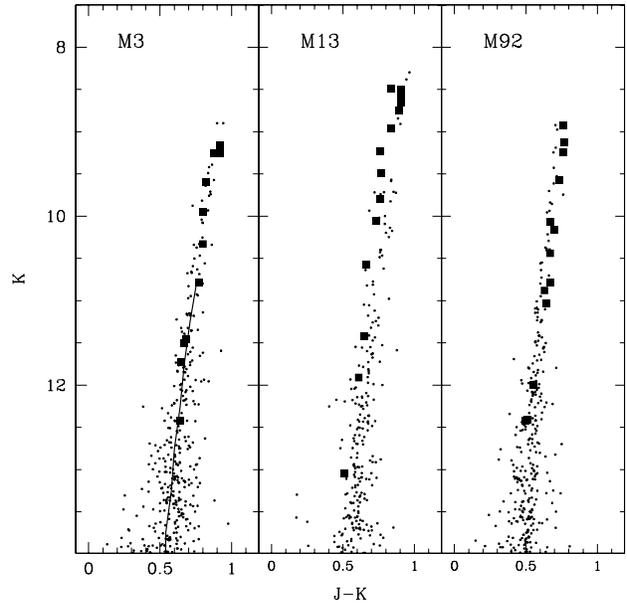}
\caption{Comparison with previous photometry for three clusters: M3, M13 and M92. The bright giants by CFP78
are overplotted to our data as filled squares. The RGB mean ridge line by \citet{L01} for M3 is also plotted.}
\label{cohen}
\end{figure}

\citet{CFP78} (hereafter CFP78) presented JHK aperture photometry for a selected sample of bright giants in the outer
region of M3, M13 and M92. The advent of a modern generation of IR arrays has
led to the study of the Turn Off (TO) region \citep[see e.g.][for M3, M13 and M92 respectively]{DH95a,DH95b,D99,L01},
or the MS 
\citep[see e.g.][for M13]{BL92}.
However, IR observations devoted to the study of the photometric properties of the RGB for the program
clusters are still lacking and a systematic star to star comparison with the quoted photometric studies is not
possible. As an example, Fig.~\ref{cohen} shows the comparison of our set with the few stars observed by
CFP78, in the outer regions of M3, M13 and M92. For M3 the RGB mean ridge line by
\citet{L01} is also shown. As can be seen the agreement with CFP78 database is acceptable, only in M13 does
our dataset appear systematically redder than the CFP78 data by ${\sim}0.03$. In M3 the agreement with the mean ridge
line of \citet{L01} is very good.

\section{The main RGB features}
In this section the main photometric indices and RGB features are derived and compared 
with those found by F00.

\subsection{The RGB fiducial ridge lines}

\begin{figure}
\centering
\includegraphics[width=8.7cm]{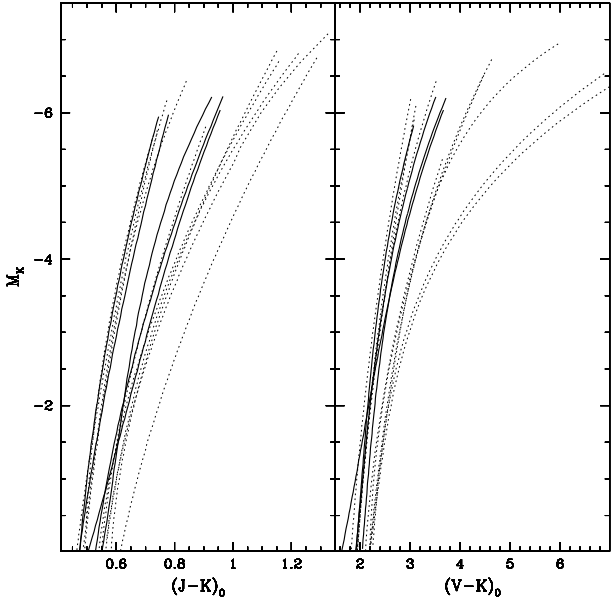} 
\caption{RGB fiducial ridge lines for the observed GGCs (solid lines) in the $M_K$,(J-K)$_0$ (left panel)
 and M$_K$,(V-K)$_0$ (right panel) compared with those of F00 (dotted lines).}
\label{line}
\end{figure}

In order to obtain the RGB fiducial ridge lines for our cluster sample we followed
the same strategy as in F00. First, we removed
the HB stars from the CMDs and then we computed the fiducial ridge lines
by using a low-order polynomial to fit the RGB stars, rejecting those at
${\geq}{\pm}2{\sigma}$ from the best-fit line. 

The ridge lines obtained following this procedure are overplotted to the (K,J--K) and (K,V--K) CMDs shown in Fig.~\ref{cmd}.
In order  to convert the RGB fiducial ridge lines into the absolute plane we adopted the distance scale defined in F99.
The correction for reddening has been computed by using the reddening values listed in Table 1 (F99) and 
the extinction coefficient for the J and K bands, reported by
\citet{SM79} $A_J/E(B-V)=0.87$ and $A_K/E(B-V)=0.38$ . 
The adopted values for the true distance modulus and
reddening are listed in Table\ref{par}.

Fig.~\ref{line} shows the observed RGB fiducial ridge lines in the absolute
$M_K$,(J-K)$_0$ and $M_K$,(V-K)$_0$ planes for the observed clusters (heavy solid lines), 
the ridge lines of the 10 GGCs presented by F00 (dotted lines) are also plotted as reference. 
As expected, the mean ridge lines of the five
intermediate--low metallicity clusters studied here lie in the bluer region of the diagrams.

\subsection{The RGB location in Color and in Magnitude}

\begin{table*}
\begin{center}
\caption[]{RGB $\rm (J-K)_0$ colors for the observed GCs at different magnitudes ($\rm M_K=-3,-4,-5,-5.5$).}
\label{Tjk}
\begin{tabular}{lccccc}
\hline\hline
\\
Name & $[M/H]$ & $(J-K)_0^{-5.5} $ & $(J-K)_0^{-5} $ &$(J-K)_0^{-4} $& $(J-K)_0^{-3}$\\
\\
\hline
 M3 & -1.16&0.84${\pm}0.02$&0.79${\pm}0.02$&0.72${\pm}0.01$&0.66${\pm}0.01$ \\
 M5 &-0.90&0.90${\pm}0.02$&0.86${\pm}0.02$&0.78${\pm}0.02$&0.71${\pm}0.01$\\
 M10 &-1.25&0.75${\pm}0.01$&0.72${\pm}0.01$&0.66${\pm}0.01$&0.60${\pm}0.01$\\
 M13& -1.18&0.89${\pm}0.02$&0.85${\pm}0.02$&0.76${\pm}0.02$&0.69${\pm}0.01$\\
 M92&-1.95&0.71 ${\pm}0.01$&0.68 ${\pm}0.01$&0.62${\pm}0.01$&0.57${\pm}0.01$\\
\hline
\end{tabular}
\end{center}
\end{table*}

\begin{table*}
\begin{center}
\caption[]{RGB $\rm(V-K)_0$ colors for the observed GCs at different magnitudes ($\rm M_K=-3,-4,-5,-5.5$).}
\label{Tvk}
\begin{tabular}{lccccc}
\hline\hline
\\
Name & $[M/H]$ & $(V-K)_0^{-5.5} $ & $(V-K)_0^{-5} $ &$(V-K)_0^{-4} $& $(V-K)_0^{-3}$\\
\\
\hline
 M3 &-1.16&3.33${\pm}0.09$&3.10${\pm}0.08$&2.73${\pm}0.06$&2.47${\pm}0.04$ \\
 M5 &-0.90&3.38${\pm}0.09$&3.15${\pm}0.09$&2.76${\pm}0.07$&2.44${\pm}0.06$   \\
 M13&-1.18&3.16${\pm}0.09$&2.96${\pm}0.07$&2.62${\pm}0.05$&2.38${\pm}0.04$\\
 M92&-1.95&2.95${\pm}0.07$&2.76${\pm}0.06$&2.50${\pm}0.05$&2.30${\pm}0.03$\\
\hline
\end{tabular}
\end{center}
\end{table*}

\begin{figure}
\centering
\includegraphics[width=8.7cm]{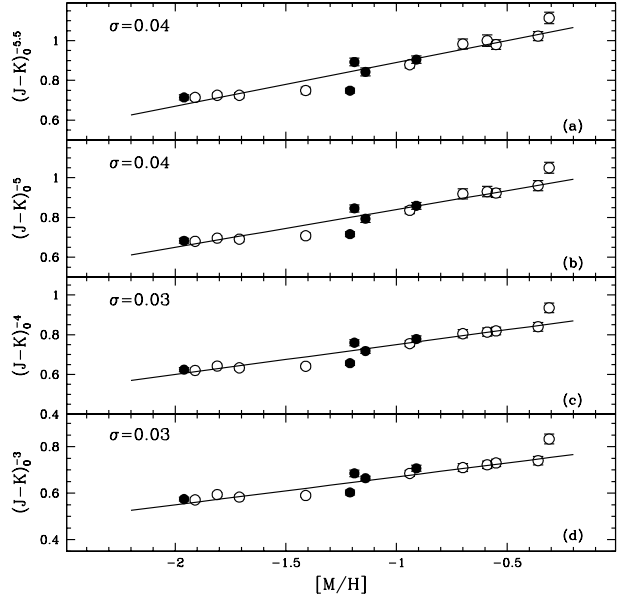}
\caption{RGB mean (J-K)$_0$ color at different magnitudes ($\rm M_K=-3,-4,-5,-5.5$) as a function of the
global metallicity for the observed clusters (filled circles). The data from F00
sample (empty circles) are also shown for comparison.
The solid lines are the best-fit to the F00 data, while ${\sigma}$ is the standard deviation of the total sample
 (F00 and this work).}
\label{coljk}
\end{figure}

\begin{figure}
\centering
\includegraphics[width=8.7cm]{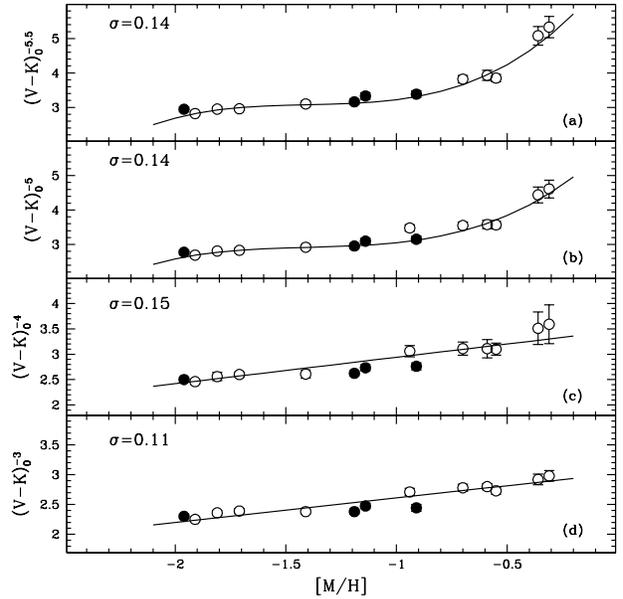}
\caption{RGB mean (V-K)$_0$ color at different magnitudes ($\rm M_K=-3,-4,-5,-5.5$) as a function of the
global metallicity for the observed clusters (filled circles) and for the F00 
sample (empty circles).
The solid lines are the best-fit to the F00 data, and ${\sigma}$ is the computed standard deviation of the total 
sample (F00 + this work).}
\label{colvk}
\end{figure}

In order to define the mean RGB properties of our cluster sample we used the observables defined
in F00, namely: {\it i)}~the intrinsic {\rm (J--K)}$_0$ and {\rm (V--K)}$_0$ colors at fixed absolute magnitude
${\rm (M_K=-3,-4,-5,-5.5)}$ and {\it ii)}~the ${\rm M_{K}}$ absolute magnitude 
at constant color, both in the ${\rm M_K}$,{\rm (J--K)}$_0$ and ${\rm M_K}$,{\rm (V--K)}$_0$ planes. 
The intrinsic {\rm (J--K)}$_0$ and {\rm (V--K)}$_0$ colors of the RGB measured at different ${\rm M_K}$
 are listed in Tables \ref{Tjk} and \ref{Tvk}, respectively. 

In order to perform a complete characterization of the RGB, 
these observables have been calibrated as a function of the
cluster global metallicity ([M/H]) defined and computed in F99. As discussed in that paper the global metallicity take in 
account the contribution of the ${\alpha}$--elements in the definition of the {\it total metallicity} of the cluster.

Fig.~\ref{coljk} and \ref{colvk} show the mean RGB {\rm (J--K)}$_0$ and {\rm (V--K)}$_0$ colors respectively,
 at different magnitude levels as a
function of the global metallicity (filled circles). The value measured by F00 in the reference
 10 GGCs are plotted as empty
circles. The best fit to the F00 data is also shown in each panel. The five additional points, from this study,
well fit into the F00 relations. Note that the Y\--scale of panel (a) and (b) of Fig.~\ref{colvk}
spans a much larger range in the V\--K color than panel (c) and (d). This
could produce the incorrect impression that the upper part of the RGB does not correlate with metallicity. Indeed, the
linear region of the relations show in panel (a) and (b) have a slope (0.71 and 0.61, respectively) significantly
 larger than
those shown in panel (c) and (d) (0.44 and 0.34, respectively). This evidence quantitatively confirms (according to 
Fig.~\ref{line}) that the location in color of the upper part of the RGB is much more sensitive to the 
metallicity than the lower
part.

Fig.~\ref{mK} 
shows the trend of the $\rm M_K$ absolute magnitudes at fixed {\rm (V--K)}$_0$=3 and {\rm (J--K)}$_0$=0.7
colors as a function of the global metallicity, 
the solid lines being the best-fit to the F00 data. The derived values for{ $\rm M_K$ at different {\rm (V--K)}$_0$ and
{\rm (J--K)}$_0$ are listed in Table \ref{mags}.
The values measured in our sample well fit into the relations calibrated by F00.

\begin{figure}
\centering
\includegraphics[width=7cm]{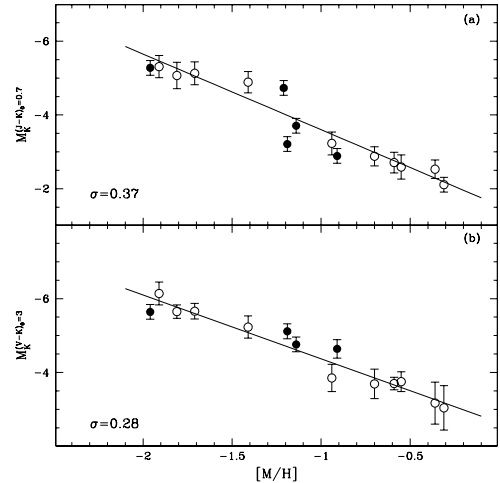}
\caption{$M_K$ at constant (J-K)$_0$=0.7 (upper panel) and (V-K)$_0$=3 (lower panel) as a
function of the global metallicity for the clusters in our sample (filled circles) and
for the F00 sample (empty circles).
The solid lines are the best-fit to the F00 data and ${\sigma}$ is the data standard deviation of the global sample
 (F00 and this work).}
\label{mK}
\end{figure}

The errors on the derived absolute $\rm M_K$ have been computed by combining the two main sources of errors: the
uncertainty in the absolute distance modulus (0.2 mag) and the uncertainty due to the reddening. In fact, given the
intrinsic steepness of the RGB, an error of a few hundredths of magnitude in the reddening correction easily implies
0.15\--0.20 mag uncertainty in the derived {\rm $M_K$} absolute magnitude, depending on the height along the RGB (see  
Fig.~\ref{line}).

\subsection{The RGB slope}

\begin{figure}
\centering
\includegraphics[width=7cm]{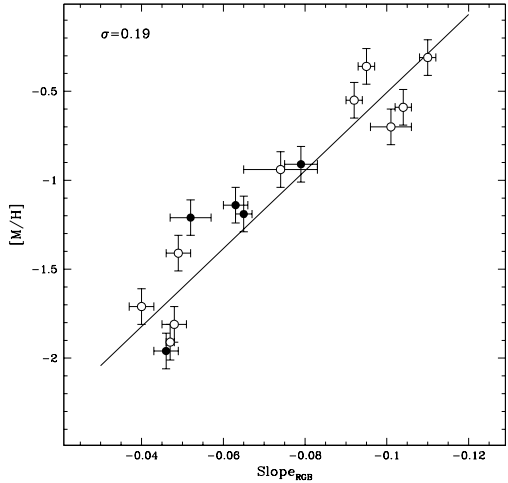}
\caption{Global metallicity as a function of the derived RGB slope for the selected 5 GGCs (filled 
circles) and for the F00 sample (empty circles).
The solid lines is the best-fit to the F00 data and ${\sigma}$ is the standard deviation of the global sample (F00 and
this work).}
\label{slope}
\end{figure}

One of the most useful parameters to derive the cluster metallicity 
is the slope of the brightest portion of the RGB (namely brighter than the HB), since
 it is a reddening and distance independent measurement.
A careful estimate of the RGB slope is a complicated task, even in the {\rm K,J--K} plane where
the RGB morphology is less curved than in any other plane. 
However, adopting the technique
described in K95 and KF95 we computed the RGB slope
by means of a linear fit. The derived values are listed in Table \ref{mags}, and Fig \ref{slope} shows
the behaviour of this parameter as a function of the global metallicity. As usual, the analogous measurements
 and the average relation
from F00 are plotted for comparison.

\subsection{The RGB bump}

\begin{table*}
\begin{center}
\caption{Calibrated RGB features for the observed clusters.}
\label{mags}
\begin{tabular}{lcccccccc}
\hline\hline
\\
Name & $[M/H]$ & $M_K^{(J-K)_0=0.7}$ & $M_K^{(V-K)_0=3}$&Slope$_{RGB}$&$K^{Bump}$&$M_K^{Bump}$&$K^{Tip}$&$M_K^{Tip}$\\
\\
\hline
M3 & -1.16&-3.71${\pm}0.22$&-4.76${\pm}0.22$&-0.063${\pm}0.003$&13.15${\pm}0.05$&-1.88${\pm}0.21$&8.90${\pm}0.09$&-6.13${\pm}0.20$\\
M5&-0.90 &-2.89${\pm}0.22$ &-4.64${\pm}0.22$ &-0.079${\pm}0.004$&12.70${\pm}0.06$&-1.68${\pm}0.24$&8.48${\pm}0.12$&-5.90${\pm}0.22$\\
M10&-1.25 &-4.73${\pm}0.22$ &------ &-0.052${\pm}0.005$&------&------&------&------\\
M13& -1.18 &-3.21${\pm}0.22$ &-5.12${\pm}0.22$ &-0.065${\pm}0.002$&12.45${\pm}0.05$&-1.99${\pm}0.21$&8.30${\pm}0.17$&-6.14${\pm}0.21$\\
M92&-1.95 &-5.28${\pm}0.22$ &-5.64${\pm}0.22$&-0.046${\pm}0.003$&12.40${\pm}0.05$&-2.39${\pm}0.21$&8.97${\pm}0.24$&-5.82${\pm}0.21$\\
\hline
\end{tabular}
\end{center}
\end{table*}

\begin{figure}
\centering
\includegraphics[width=7cm]{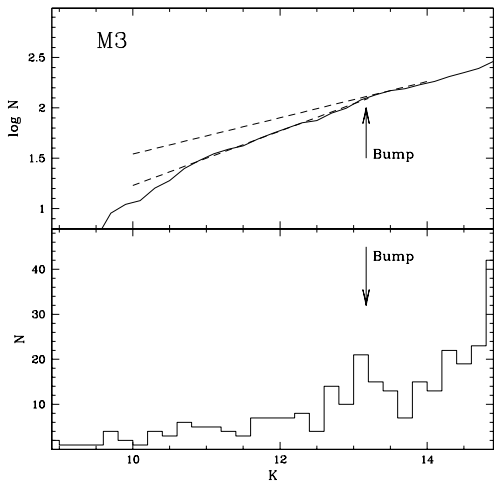}
\caption{Observed integrated (upper panel) 
and differential (lower panel) LF for M3. The dashed lines, in the upper panel, are the linear fits to the regions
above and below the RGB bump.}
\label{m3LF}
\end{figure}

\begin{figure}
\centering
\includegraphics[width=8.7cm]{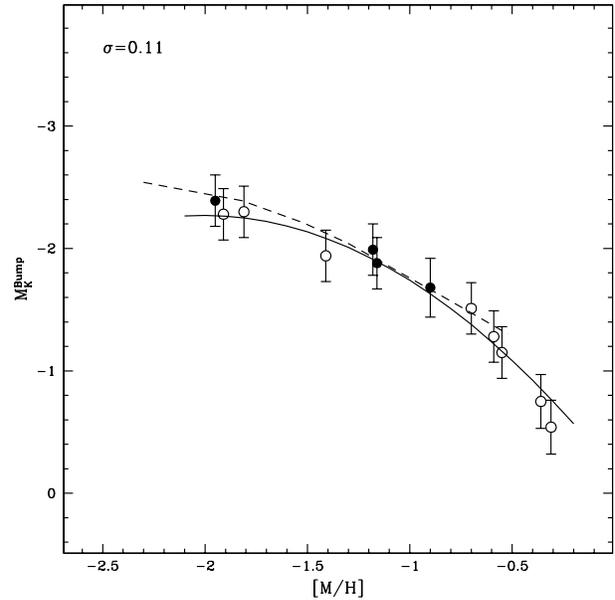}
\caption{Absolute K magnitude at the RGB bump as a function of the global metallicity for our sample
(filled circles) and for F00 dataset (empty circles). The solid line is the best fit to the F00 data. The
dashed line is the theoretical prediction by \citet{S97}.}
\label{bump}
\end{figure}

\begin{figure}
\centering
\includegraphics[width=8.7cm]{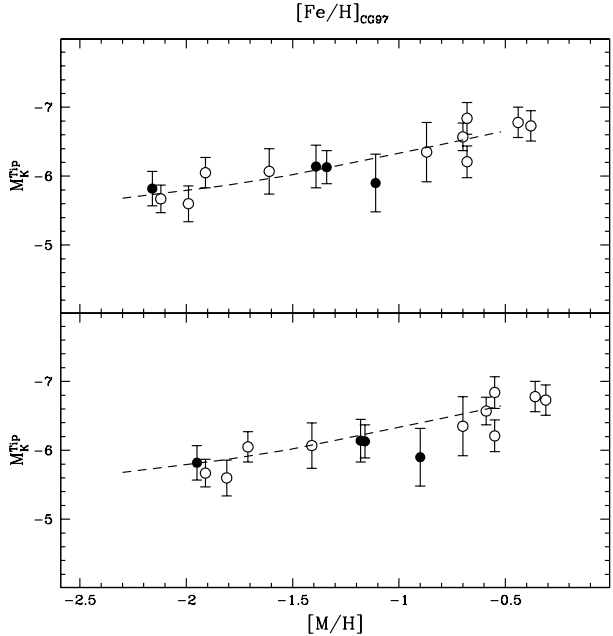}
\caption{Absolute K magnitude of the brightest star (RGB tip) as a function of the metallicity in the CG97 (upper
panel) and global scale (lower panel) for the GGCs in our sample
(filled circles) and for F00 data--set (empty circles). The dashed lines are the theoretical expectations based
on \citet{S97} models.}
\label{tip}
\end{figure}

The RGB bump is an important evolutionary feature, since it flags the point (during the post--MS evolution of low mass 
stars)
when the narrow hydrogen--burning shell reaches the discontinuity in the hydrogen distribution profile, generated  by the
previous innermost penetration of the convective envelope. 
Besides the obvious check of the accuracy of the theoretical models of stellar evolution, the identification of this 
feature
in stellar aggregates can be used as a useful tool to provide observational constraints on a number 
of population
parameters, since the RGB bump is a sensitive function of the metal content, helium abundance and the stellar population
age.
From an observational point of view, it was identified for the first time by
 \citet{KDD85} in 47 Tuc. Then, during the last 10 years this feature was identified in a large number of GGCs
\citep[see e.g.][F99 and F00]{Fu90,Zoc00}
and in other stellar aggregates \citep[see e.g.][]{Bel01,Bel02,Mon02}. 
 As emphasized by \citet{Fu90} and by F99, the combined use of the differential and integrated LFs is the best
 tool in order
to properly detect this feature.
The bump luminosity increases with decreasing metallicity, hence its identification is more difficult in
metal--poor clusters, where the brightest portion of the RGB is poorly populated because of the high
evolutionary rate of stars at the end of the RGB evolution.

Following the \citet{Fu90} prescription, we identified the RGB bump in
M3, M5, M13 and M92 by using
the integral and differential LFs in the K band (as an example, both LFs are shown in Fig.~\ref{m3LF} for M3).
Since for M10 we did not map the central (more populated) regions of the cluster, the LF is much less populated 
than those
observed in the other program clusters, preventing any reliable detection of the RGB bump. Note that the region
covered by our observations allows to sample a significant fraction of the cluster light (ranging from 25\% to 39\% in 3 out
of the 5 clusters presented here). However, in order to further increase the fraction of the sampled cluster population we
also used complementary data from 2MASS catalog (after the application of the transformations quoted in \S2). The level of
the RGB\--bump derived from our data was re\--computed on the combined sample. Also, the RGB\--bump magnitude in the K band
was compared (by using the V\--K color) with the RGB\--bump in the V magnitude from F99
finding a very good agreement. 
The observed K$^{Bump}$ and absolute 
M$_K^{Bump}$ magnitude for the four clusters, where the bump has been identified, 
are listed in Table \ref{mags}, while Fig.~\ref{bump} shows the
 absolute K magnitude of the bump as a function of the global metallicity. 
The estimates for this additional sample of clusters well fit into the empirical relation by F00, with a total data
standard deviation of ${\sigma}=0.11$.
Fig.~\ref{bump} also shows an excellent agreement with the theoretical
expectations based on the \citet{S97} models  \citep[for an age of $t=14$ Gyr, see e.g.][]{Gra03}. 

\subsection{RGB Tip}

The luminosity of the RGB tip (the bright end of the RGB) flags the end of the evolution along this sequence. 
The RGB tip
luminosity is now recognized as a valuable standard candle \citep[see e.g][ and references therein]{Bel01} 
and it has been
widely used to derive distance to extragalactic stellar populations \citep[see e.g.][ and references therein]
{fer00a,fer00b}.
A well\--defined relation between the bolometric luminosity of the brightest star and the cluster metallicity
has been found by \citet{fro81} and FCP83.
F00 presented a more recent calibration of the RGB
tip both in the K and bolometric magnitudes.

Here we present an estimate of the RGB tip for the program clusters by using the brightest giant in our observed sample
(see F00). In principle,
the magnitude of the brightest RGB star should give a reasonable estimate of the RGB tip if a significant
 fraction of 
the cluster has been sampled. Since we mapped,
in most of the program clusters, the very central (more populated) regions we can reasonable apply this technique to our
sample.  
The uncertainty of the procedure is discussed in F00, here we just note that the
possible contamination of bright AGB stars is low since these stars are significantly less than the RGB stars, and 
especially in the intermediate--low metallicity clusters, no long\--period AGB variables are expected. 
The main source of uncertainty
is surely due to statistical fluctuations; following the prescription of F00 we estimated the error expected on the basis of
the number of stars in the brightest two magnitude bin along the RGB. In our cluster sample ${\sigma}_{stat}$ turns to be
${\sim}$0.15 and we assume this as the main source of error in the RGB tip determination. The
observed K magnitude of the brightest star and its absolute magnitude M$_K$ are listed in columns 8 and 9, respectively,
of Table \ref{mags}. Fig.~\ref{tip} shows the absolute K magnitude of the tip as a function of the metallicity in the CG97 
and global scales, respectively. 
The theoretical expectation from \citet{S97} models for $t=14$ Gyr is also overplotted to the
data. The data from F00 are also shown as empty circles. As can be seen from the
figure, the theoretical prediction nicely agrees with the observations further suggesting that the adopted distance scale,
from F99, is not affected by large systematic errors.

\section{Conclusions}

Using near-IR TNG observations of five low--metallicity GCCs, a detailed analysis of the RGB features has been performed. 
From the study of CMDs and LFs we derive and calibrate severals observables describing the 
morphology and the chemical properties of the RGB and the major RGB evolutionary features, namely 
\begin{itemize}
\item the location in color and in magnitude of the RGB in the
K,J--K and K,V--K planes 
\item the RGB slope 
\item the K--band absolute magnitude of the RGB bump and tip.
\end{itemize}
These observable have been also reported in the absolute planes by adopting the metallicity and distance scales 
defined in F99 and F00.
The clusters presented in this paper represent an extension of the F00 sample, in the intermediate metal--poor regime. In a
forthcoming paper an additional sample of intermediate--high metallicity clusters will be presented 
and updated relations for all the
parameters will be derived and discussed.
\\
\\
\\
We warmly thank Paolo Montegriffo for assistance during the catalog crosscorrelation procedure and the TNG staff for
assistance during the observations.
The financial support of
the Agenzia Spaziale Italiana and the Ministero della Istruzione e della Ricerca Universitaria is kindly acknowledged.

This publication 
makes use of data products from the Two Micron All Sky Survey, which is a joint
project of the University of Massachusetts and the Infrared Processing 
and Analysis Center/California Institute
of Technology, founded by the National Aeronautics and Space 
Administration and the National Science Foundation.

\end{document}